\documentstyle[aps,twocolumn,epsfig]{revtex}
\begin{document}
\draft
\twocolumn[\hsize\textwidth\columnwidth\hsize\csname@twocolumnfalse\endcsname
\title{Multipartite entangled states in coupled quantum dots and cavity-QED}
\author{Xiaoguang Wang$^{1,2}$, Mang Feng$^{2}$, and Barry C Sanders$^{1}$}
\address{1. Department of Physics and Centre for Advanced Computing -- Algorithms and Cryptography, \\
Macquarie University, Sydney, New South Wales 2109, Australia}
\address{2. Institute for Scientific Interchange (ISI) Foundation,\\
 Viale Settimio Severo 65, I-10133 Torino, Italy}
\date{\today}
\maketitle

\begin{abstract}
We investigate the generation of multipartite entangled state in a system of $N$ quantum dots embedded in a microcavity and examine the emergence of genuine multipartite entanglement by three different characterizations of entanglement. At certain times of dynamical evolution one can generate multipartite entangled coherent exciton states or multiqubit $W$ states by initially preparing the cavity field in a superposition of coherent states or the Fock state with one photon, respectively. Finally we study environmental effects on  multipartite entanglement generation and find that the decay rate for the entanglement is proportional to the number of excitons.
\end{abstract}

]
\section{Introduction}

Quantum information processing offers important capabilities for quantum 
communications and quantum computation in a variety of physical systems, and the solid-state implementation is one of the most promising candidates. 
Semiconductor quantum dots (QDs) are appealing for the
realization of quantum computer and quantum information processing. Spatial confinement of excitons in three dimensions leads to discrete
energy levels, and the main causes of phase decoherence, namely scattering events, are highly suppressed \cite{Deco}. In this respect, QDs are very promising compared to other semiconductor structures with band states. Moreover a quantum computer scheme based on quantum dot arrays is scalable up to $\geq
100$ qubits.

Entanglement is a essential source for quantum computing and quantum information processing. It is well-known that a controlled-NOT gate can generate a maximally entangled state. Various schemes for the realization of controlled-NOT gates in QDs \cite {Gate,Loss,Imamoglu} are proposed with either the electron spin \cite{Loss,Imamoglu,Pazy} or
the discrete electronic charge degree of freedom as qubits\cite
{Biolatti,XinQi,Goan}. 
For example, Imamo\={g}lu {\it et al.} \cite{Imamoglu} proposed a
scheme that realized a controlled-NOT gate 
$C_{\text{NOT}}^{ij}$ 
between two distant quantum dots $i$ and $j$ via the cavity quantum electromagnetic dynamics (CQED) techniques. 
Here the first superscript $i$ in $C_{\text{NOT}}^{ij}$ denotes the control and the second denotes the target. These controlled-NOT gates in turn can be used to generate a tripartite maximally entangled Greenberger-Horne-Zeilinger (GHZ)\cite{GHZ} state as $|0\rangle\otimes|0\rangle\otimes|0\rangle+|1\rangle\otimes|1\rangle\otimes|1\rangle=C_{\text{NOT}}^{13} C_{\text{NOT}}^{12}
(|0\rangle+|1\rangle)\otimes |0\rangle\otimes |0\rangle.$ 
Quiroga {\it et al}. showed \cite{Luis}  that an optically
controlled exciton transfer process can generate maximally
entangled Bell states \cite{Bell} and GHZ states. To demonstrate that excitons in coupled quantum dots are ideal candidates for reliable preparation of entangled states in solid-state systems, Chen {\em et al}.\cite{Chen} experimentally 
investigated how to optically induce entanglement of excitons in a single gallium arsenide QD: the entanglement is identified by the spectrum of the
phase-sensitive homodyne-detected coherent nonlinear optical response.

Although no experimental observation of entanglement of many-exciton systems has yet occurred, the theoretical studies are necessary if we desire to have a large-scale quantum computing device. In a recent investigation by 
Liu $et~al$ \cite{Yuxi}{\it \ } for generating bipartite entangled coherent exciton states in a system of two coupled quantum dots
and CQED with dilute excitons, they found that the bipartite maximally entangled coherent exciton 
states can be generated
when the initial cavity field is in an odd coherent state. The relation between entanglement of exciton and photon numbers in the cavity was also discussed. 

It is interesting to extend to  multipartite entanglement in such systems. Multipartite entanglement is not only of intrinsic interest itself but also of practical importance in quantum information proposals such as quantum teleportation and quantum cryptography.
One motivation to consider multipartite entanglement in many QDs is that around 10 QDs can be embedded in a microdisk structure and coupled to a single cavity mode in the quantum information process scheme proposed by Imamo\={g}lu {\it et al.}
We will first extend the model of Ref. [11] to the case of many QDs , and
then study how to generate multipartite entangled states and examine multipartite entanglement in such systems.

The paper is organized as follows. In Sec.\,II, we present a model for 
coupled QDs in CQED and determine the exact state vector at any time $t$, which will be shown to be a multipartite entangled state in some time ranges. 
In Sec.\,III, we use three different methods to examine the multipartite entanglement in the state. Then in Sec.\,IV, we show that multipartite entangled coherent exciton states can be generated at certain times during the dynamical evolution. 
In Sec. V, we analyse the effects of environment on the generation of multipartite entanglement. The conclusions are given in Sec.\,VI.

\section{Multipartite entangled states}

We consider quantum dots embedded in a single-mode cavity. We assume that
the QDs are sufficiently large and there are only a few electrons excited from valence band to conduction band \cite{Yuxi}. 
In the assumption of low exciton excitation density, exciton operators can be approximated by
boson operators, and all nonlinear dynamics such as exciton-exciton
interaction can be neglected. As the distance between any two excitons is also
assumed to be large, the interaction between any two excitons can be safely 
neglected. There exists a resonant interaction between the excitons and the
cavity field. The Hamiltonian under the rotating wave approximation is given
by ($\hbar =1$)\cite{Yuxi,Model}

\begin{equation}
H=\omega b_0^{\dagger }b_0+\omega \sum_{n=1}^Nb_n^{\dagger
}b_n+\sum_{n=1}^Ng_n(b_n^{\dagger }b_0+b_0^{\dagger }b_n),
\end{equation}
where $b_0^{\dagger }$ $(b_0)$ is the creation (annihilation) operator of the
cavity field with frequency $\omega ,$ $b_n^{\dagger }(b_n)$ ($n=1,2,\cdots,N$)
denote the creation (annihilation) operator of the $n^{\text{th}}$ exciton with the same
frequency $\omega ,$ $g_n$ is the coupling constant between the cavity
field and $n^{\text{th}}$ exciton, and $N$ is the number of excitons.

It is convenient to write the coupling constants $g_i$ in generalized
spherical coordinates \cite{Recati,Spherical} as
\begin{eqnarray*}
g_1 &=&G\cos \phi _1, \\
g_2 &=&G\sin \phi _1\cos \phi _2, \\
g_3 &=&G\sin \phi _1\sin \phi _2\cos \phi _3, \\
&&\vdots \\
g_{N-1} &=&G\sin \phi _1\sin \phi _2\cdots \cos \phi _{N-1}, \\
g_N &=&G\sin \phi _1\sin \phi _2\cdots \sin \phi _{N-1},
\end{eqnarray*}
where $G=\sqrt{\sum_{k=1}^Ng_k{}^2}.$ The evolution operator corresponding
to the Hamiltonian is then obtained as

\begin{eqnarray}
U(t) &=&V_{N-1,N}^{\dagger }(\phi _{N-1})\cdots 
V_{2,3}^{\dagger }(\phi
_2)
V_{1,2}^{\dagger }(\phi _1)
\tilde{V}_{0,1}(t)V_{1,2}(\phi _1)  \nonumber \\
&&\times V_{2,3}(\phi _2)\cdots V_{N-1,N}(\phi _{N-1})U_0(t),
\end{eqnarray}
where
\begin{eqnarray}
V_{i,j}(\phi _i) &=&\exp [\phi _i(b_i^{\dagger }b_j-b_j^{\dagger }b_i)]\text{ 
}(i\neq j),  \nonumber \\
\tilde{V}_{0,1}(t) &=&\exp [-iGt(b_0^{\dagger }b_1+b_1^{\dagger }b_0)], 
\nonumber \\
U_0(t) &=&\exp [-i\omega t(b_0^{\dagger }b_0+\sum_{i=1}^Nb_i^{\dagger }b_i)].
\end{eqnarray}
Both $V_{i,j}(\phi _i)$ and $\tilde{V}_{0,1}(t)$ are operators for the beam
splitters. We see that the evolution operator can be written as a free
evolution $U_0(t)$ followed by a series of beam splitter operators.

Let us assume that the initial state is $|\psi (0)\rangle =|\alpha \rangle
_{_0}\otimes |0\rangle _{_1}\otimes \cdots \otimes |0\rangle _{_N},$ i.e.,
the cavity field is in a coherent state $|\alpha \rangle _{_0}$ with $\alpha\neq 0$ and all
the excitons are in  vacuum states. After the action of the unitary
operator $U(t) ,$ the state evolves into 
\begin{equation}
|\psi (t)\rangle=|\alpha_0(t),\alpha_1(t),\cdots,\alpha_N(t)\rangle_{0\cdots N},
\end{equation}
where $
|\alpha_0(t),\cdots,\alpha_N(t)\rangle_{0\cdots N}
\equiv|\alpha _0(t)\rangle _0\otimes |\alpha _1(t)\rangle
_1\otimes \cdots \otimes |\alpha _N(t)\rangle _N$ and 
\begin{eqnarray}
\alpha _0(t) &=&\alpha \cos (Gt)e^{-i\omega t},  \nonumber \\
\alpha _n(t) &=&-i\alpha \sin (Gt)g_n/Ge^{-i\omega t},\, n=1,2,\cdots,N.  
\label{eq:coefs}
\end{eqnarray}
Thus, if the initial state of the cavity field is in a superposition
of coherent states, the resulting state will be an entangled coherent state.
Let us assume the initial state of the cavity field to be in a superposition of
two coherent states $|\pm \alpha \rangle $ and the excitons to  be in vacuum states, i.e.,
\begin{eqnarray}
|\Psi (0)\rangle &=&[2+2\cos \theta \exp (-2|\alpha |^2)]^{-1/2}  \nonumber
\\
&&(|\alpha \rangle _{_0}+e^{i\theta }|-\alpha \rangle _{_0})\otimes
|0\rangle _{_1}\otimes \cdots \otimes |0\rangle _{_N}.  \label{eq:initial}
\end{eqnarray}
Specifically, for $\theta =0,$ $\pi ,$ and $\pi /2,$ the cavity
superposition state reduces to even, odd\cite{Buzek}, and Yurke-Stoler%
\cite{YS} coherent states, respectively. Then the state vector at time $t$ is given by
\begin{eqnarray}
|\Psi(t)\rangle&=&
{\cal N}(|\alpha_0(t),\alpha_1(t),\cdots,\alpha_N(t)\rangle_{0\cdots N}\nonumber\\
&&+e^{i\theta}|-\alpha_0(t),-\alpha_1(t),\cdots,-\alpha_N(t)\rangle_{0\cdots N}),\label{eq:state}
\end{eqnarray}
where  
\begin{equation}
{\cal N}=[2+2\cos \theta
\prod_{k=0}^Np_k(t)]^{-1/2}
\end{equation}
is the normalization constant, and $p_k(t)\equiv\exp (-2|\alpha _k(t)|^2)$ is the overlap of the two coherent states $|\pm \alpha _k(t)\rangle .$ The resulting state $%
|\Psi (t)\rangle $ is a multipartite entangled coherent state\cite
{ECS,WangBarry}.

We now choose the orthogonal
basis\cite{WangBarry} 
\begin{equation}
|{\bf 0}\rangle _k\equiv |\alpha _k(t)\rangle ,\text{ }|{\bf 1}\rangle _k\equiv
(|-\alpha _k\rangle -p_k(t)|{\bf 0}\rangle )/{\cal M}_k(t),  \label{eq:ortho}
\end{equation}
where ${\cal M}_k(t)=\sqrt{1-p_k(t)^2}.\,$ It then follows that 
\begin{equation}
|-\alpha _k(t)\rangle
={\cal M}_k(t)|{\bf 1}\rangle +p_k(t)|{\bf 0}\rangle.
\end{equation} 
Using this basis the state vector at time $t$ can be rewritten as
\begin{eqnarray}
|\Psi (t)\rangle &=&{\cal N}\{|{\bf 0}\rangle _0\otimes\cdots \otimes |{\bf 0}\rangle _N  
+e^{i\theta}
[{\cal M}_0(t)|{\bf 1}\rangle_0 +p_0(t)|{\bf 0}\rangle_0]\nonumber\\
&&\otimes\cdots\otimes
[{\cal M}_N(t)|{\bf 1}\rangle_N +p_N(t)|{\bf 0}\rangle_N]
\}.
\label{eq:qubit}
\end{eqnarray}
After the `encoding' this state is a multiqubit state. Then we can fully exploit the sophisticated tools available for examining multipartite entanglement of qubits to study our state and determine if it is genuinely multipartite entangled. 

\section{Examination of multipartite entanglement}

We study the multipartite entanglement by examining 
(i) Mermin-Klyshko inequality\cite{Mermin,Klyshko}, (ii) state preparation fidelity \cite{Sackett,Uffink}, and (iii) the square of the multiqubit concurrence\cite{Cof00,Won01}.
The first two,  namely the Mermin-Klyshko inequality and the state preparation fidelity, are related to two sufficient conditions that distinguish between genuinely $N$-partite entangled states and those in which only $M$ particle are entangled ($M<N$). 
The last one, the square of the multiqubit concurrence, is not by itself a measure of $N$-particle entanglement, but it appears to be related to some kind of multipartite entanglement. That is, we can gain some information about the degree of multipartite entanglement by calculating the square of the multiqubit concurrence. 

\subsection{Mermin-Klyshko inequality}

Let us first use the Mermin-Klyshko inequality\cite{Mermin,Klyshko} to examine the $N$-partite entanglement. This inequality generalizes the Bell inequality\cite{Bell} and Clauser-Horne-Shimony-Holt (CHSH) inequality\cite{CHSH}, which not only tests the predictions of quantum mechanics against those of local hidden variable theory but also distinguishes entangled from non-entangled states\cite{Gisin,Uffink}. 
The Mermin-Klyshko inequality is\cite{Mermin,Klyshko} 
\begin{equation}
|\langle {\cal B}_N\rangle|\le 2,
\end{equation}
where ${\cal B}_N $ is the Bell operator defined recursively as
\begin{equation}
{\cal B}_N =\frac{1}{2}(A_N+A'_N)\otimes {\cal B}_{N-1}+\frac{1}{2}(A_N-A'_N)\otimes {\cal B}'_{N-1},
\end{equation}
${\cal B}'_N$ is obtained from ${\cal B}_N$ by exchanging primed and unprimed terms,
\begin{equation}
{\cal B}_N^\prime =\frac{1}{2}(A_N+A'_N)\otimes {\cal B}_{N-1}^\prime-\frac{1}{2}(A_N-A'_N)\otimes {\cal B}_{N-1},
\end{equation}
${\cal B}_1=2A_1$, and ${\cal B}_1^\prime=2A'_1$. All $A_i$ and $A'_i$ are dichotomous observables. Let $S_n$ denote 
the set of all $N$-particle states and $S_N^{N-1}$ denote the subset of those states which are at most $(N-1)$-partite entangled. Then, from the results of 
Refs.\,\cite{Gisin,Uffink}, for a state $\rho$ we have
\begin{eqnarray}
&&\forall \rho \in S_N^{N-1}:|\langle {\cal B}_N \rangle| \leq 2^{N/2},\label{eq:bell1}\\
&&\forall \rho \in S_N: |\langle {\cal B}_N \rangle| \leq 2^{(N+1)/2},
\end{eqnarray}
which implies that a sufficient condition for $N$-partite entanglement 
is the violation of the inequality given by Eq.\,(\ref{eq:bell1}). Now we define a quantity $B$
\begin{equation}
B(\rho)=\frac{|\langle {\cal B}_N \rangle|-2^{N/2}}{2^{(N+1)/2}-2^{N/2}}.
\end{equation} 
Then the state $\rho$ is $N$-partite entangled when $B(\rho)>0$ and maximally entangled when $B(\rho)=1$.

Let $A_i=\sigma_x$ and $A'_i=\sigma_y$ for any $i$. Then the Bell operator ${\cal B}_N$ and ${\cal B}^\prime_N $ become\cite{Dur}
\begin{eqnarray}
{\cal B}_N&=&2^{(N+1)/2}(e^{-i\beta_N}\sigma_+^{\otimes N}+ e^{i\beta_N}\sigma_-^{\otimes N}),\nonumber\\
{\cal B}^\prime_N&=&2^{(N+1)/2}(-ie^{-i\beta_N}\sigma_+^{\otimes N}+i e^{i\beta_N}\sigma_-^{\otimes N}),
\end{eqnarray}
where $\beta_N=\pi/4(N-1), \sigma_+=|{\bf 0}\rangle\langle {\bf 1}|,$ and $\sigma_-=|{\bf 1}\rangle\langle {\bf 0}|$. For our state $|\Psi(t)\rangle$ (\ref{eq:qubit}) we have
\begin{eqnarray}
&&{\cal B} (|\Psi(t)\rangle\langle\Psi(t)|)\nonumber\\
&=&\frac{2^{(N+1)/2}}{1+\cos\theta e^{-2|\alpha|^2}}
[\cos(\theta-\beta_N)+ \cos(\beta_N)e^{-2|\alpha|^2}]\nonumber\\
&& \times \prod_{k=0}^N\sqrt{1-p_k^2}.
\end{eqnarray}
We choose $\theta=\beta_N$. Then the above equation reduces to
\begin{equation}
{\cal B}=2^{(N+1)/2}\prod_{k=0}^N\sqrt{1-p_k^2}.\label{eq:bbbbb}
\end{equation}
Therefore this sufficient condition becomes 
\begin{equation}
\prod_{k=0}^N\sqrt{1-p_k^2}>1/\sqrt{2}.
\end{equation}
Numerical results are provided later. Next we discuss another method based on  state preparation fidelity to examine multipartite entanglement.

\subsection{State preparation fidelity}

The so-called state preparation fidelity ${\cal F}$ of an $N$-qubit state $\rho$ is defined as
\begin{equation}
{\cal F}(\rho)=\langle\psi_{\text{GHZ}}| \rho|\psi_{\text{GHZ}}\rangle,
\label{eq:ff}
\end{equation}
where 
\begin{equation}
|\psi_{\text{GHZ}}\rangle=\frac{1}{\sqrt{2}}(
|{\bf 0}\rangle\otimes...\otimes|{\bf 0}\rangle+e^{i\gamma}
|{\bf 1}\rangle\otimes...\otimes|{\bf 1}\rangle)
\end{equation}
is the GHZ state.
A sufficient condition for $N$-partite entanglement is given by\cite{Sackett,Uffink}
\begin{equation}
{\cal F} (\rho)>1/2. \label{suf}
\end{equation}

{}From Eqs.\,(\ref{eq:ff}) and (\ref{eq:qubit}) the state preparation fidelity for 
the multipartite state $|\Psi(t)\rangle$ is obtained as
\begin{eqnarray}
&&{\cal F} (|\Psi(t)\rangle\langle\Psi(t)|)\nonumber\\
&&=\frac{1}{4+4\cos\theta e^{-2|\alpha|^2}} \Big[1+\prod_{k=0}^{N}(1-p_k^2)+e^{-4|\alpha|^2}\nonumber\\
&&+2\cos\theta e^{-2|\alpha|^2}
+2\cos(\theta-\gamma)\prod_{k=0}^{N}\sqrt{1-p_k^2}\nonumber\\
&&+2\cos\gamma e^{-2|\alpha|^2}\prod_{k=0}^{N}\sqrt{1-p_k^2}\Big].
\end{eqnarray}
Let $\theta=\gamma=\pi/2$. The above equation then reduces to
\begin{equation}
{\cal F}=\left(1+\prod_{k=0}^{N}\sqrt{1-p_k^2}\right)^2/4+e^{-4|\alpha|^2}/4, \label{eq:fffff}
\end{equation}
and the sufficient condition (\ref{suf}) becomes 
\begin{equation}
\left(1+\prod_{k=0}^{N}\sqrt{1-p_k^2}\right)^2+e^{-4|\alpha|^2}>2. 
\end{equation}
For convenience we define 
\begin{equation}
F=2{\cal F}-1.
\end{equation}
When $F>0$ the state is $(N+1)$-partite entangled. 

In Fig.\,1 we plot the quantity $B$ and $F$ against time $t$. 
Henceforth we assume $g_1=g_2=...=g_N$ and $G=1$.
The period of $B$ and $F$ with respect to time is $2\pi. $
We can clearly see the two sufficient conditions can be satisfied in some time ranges which means that the state is a genuine multipartite entangled state. Moreover, we find the time range in which the sufficient condition is satisfied based on the state preparation fidelity is larger than that based on the Mermin-Klyshko inequality. 

\subsection{Square of the multiqubit concurrence}
Recently Coffman {\it et al}~\cite{Cof00} used concurrence \cite{Conc} to
examine three-qubit systems, and quantified the amount of tripartite entanglement in three-qubit systems by the quantity $\tau _{0,1,2}$\cite{Cof00}
\begin{equation}
\tau _{0,1,2}=C_{0(12)}^2-C_{01}^2-C_{02}^2,  \label{eq:tau012}
\end{equation}
where $C_{0(12)}$ denotes the concurrence between qubit 1 and qubits 2 and 3.
Applying the general result for concurrence of bipartite nonorthogonal pure
states \cite{Wang} to the state (\ref{eq:qubit}) for $N=2$ yields 
\begin{equation}
C_{0(12)}=\frac{\sqrt{(1-p_0^2)(1-p_1^2p_2^2)}}{1+p_0p_1p_2\cos \theta }.
\label{eq:c012}
\end{equation} 
By the standard method for calculating the concurrence \cite{Conc,Wang}, we find
\begin{eqnarray}
C_{01} &=&\frac{p_2\sqrt{(1-p_0^2)(1-p_1^2)}}{1+\cos \theta p_0p_1p_2},\text{
}  \nonumber \\
C_{02} &=&\frac{p_1\sqrt{(1-p_0^2)(1-p_2^2)}}{1+\cos \theta p_0p_1p_2}.
\label{eq:c01}
\end{eqnarray}
The concurrence $C_{02}$ is obtained by a transformation $%
1\longleftrightarrow 2$ in the expression of $C_{01}$.
From Eqs.\,(\ref{eq:tau012}), (\ref{eq:c012}), and (\ref{eq:c01}), we obtain
\begin{equation}
\tau _{0,1,2}=\frac{\prod_{n=0}^2(1-p_n^2)}{(1+\cos \theta e^{-2|\alpha|^2})^2}.
\label{eq:tautau}
\end{equation}

Now consider the case of odd $N$ excitons ($N>2$). Wong and Christensen~ \cite{Won01} proposed the square of the multiqubit concurrence as a {\em potential measure} of the multipartite entanglement for an even-number pure
qubit state. The $(N+1)$-qubit concurrence for a  pure state $%
|\psi \rangle $ is defined as\cite{Won01}
\begin{equation}
C_{0,1,\ldots ,N}\equiv |\langle \psi |\sigma _y^{\otimes N+1}|\psi
^{*}\rangle |,  \label{eq:m1}
\end{equation}
where $\sigma _y=-i(|{\bf 0}\rangle \langle {\bf 1}|-|{\bf 1}\rangle \langle 
{\bf 0}|)$ is a Pauli matrix.

Then applying Eq.\,(\ref{eq:m1}) to the state (\ref{eq:qubit})
leads to the square of the multiqubit concurrence
\begin{equation}
\tau_{0,1,\ldots ,N}=C^2_{0,1,\ldots,N}=\frac{\prod_{n=0}^N(1-p_n^2)}{(1+\cos \theta
e^{-2|\alpha |^2})^2}.  \label{eq:tau4}
\end{equation}
From the expression of the square of multiqubit concurrence we can see that the multiqubit concurrence is unchanged by permutation of qubits, which implies that it really represents 
$(N+1)$-partite entanglement.
We note that although the above formula is obtained for odd $N,$ by
comparing (\ref{eq:tau4}) and (\ref{eq:tautau}), it is also applicable to $%
N=2.$ We also note that Eqs.\,(\ref{eq:tau4}) and (\ref{eq:tautau}) are
applicable to the more general state $(|\psi _0\rangle \otimes |\psi _1\rangle
\otimes \cdots \otimes |\psi _N\rangle 
+e^{i\theta }|\phi _0\rangle \otimes |\phi _1\rangle \otimes \cdots
\otimes |\phi _N\rangle )$
with real overlap $\langle \psi _i|\phi _i\rangle (i=0,1,...,N)$ up to a normalization constant.

In Fig.\,2 we plot the square of the multiqubit concurrence against time for $N=2, N=3,$ and $N=5.$ We observe  that the entanglement periodically reaches its maximum twice per period, and  the multipartite entanglement is suppressed with the increase of the number of excitons. One way to overcome this suppression is to increase the parameter $|\alpha|^2$. {}From Eq.\,(\ref{eq:tau4}) we know that the larger the parameter $|\alpha|^2$, the larger the multipartite entanglement.

{}From Eq.\,(\ref{eq:tau4}) we know that the square of the $(N+1)$-qubit concurrence
reaches a maximum value at $\theta=\pi$ when other parameters are fixed. This implies that the best input state is the odd coherent state in order to generate 
multipartite entanglement. On the other hand we know that the average photon number of the cavity field in the initial state $|\Psi(0)\rangle$ is
\begin{equation}
\langle b_0^\dagger b_0\rangle=\frac{1-\cos\theta e^{-2|\alpha|^2}}{1+\cos\theta e^{-2|\alpha|^2}}|\alpha|^2.
\end{equation}
We see that the mean photon number of the cavity field reaches its maximum when the field is in the odd coherent state ($\theta=\pi$) when other parameters are fixed. The mean photon number represents the energy of the system. Therefore, it turns out that the more energy contained in the initial cavity field, the larger the multipartite entanglement.

By three different methods we have examined multipartite entanglement and found that the state (\ref{eq:qubit}) is a genuine multipartite 
entangled state over a large range of parameters by each of the indicators.

\section{Multipartite entangled coherent exciton states}

It is clear that the cavity field and excitons are decoupled
when $t=n\pi /(2g)$ for integer $n.$ 
Without loss of generality we examine the state vector $|\Psi(t)\rangle$ at times $\pi $ and $%
\pi /2$ in one period. From the discussion of the last section, at these times
there is no global multipartite entanglement. At
time $t=\pi ,$ the state vector returns to the initial state (\ref
{eq:initial}) with $\alpha \rightarrow $ $-\alpha e^{-i\omega \pi },$ 
and the excitons  are in the vacuum state. It is interesting to see
that at time $t=\pi /2$ the cavity field is decoupled from the excitons; however
the excitons are left in a multipartite entangled coherent exciton states given by

\begin{eqnarray}
|\Psi \rangle _{\text{excitons}} &=&{\cal N}^{\prime}
(|\beta,\beta,\cdots,\beta\rangle_{1\cdots N}\nonumber\\
&&+e^{i\theta}|-\beta,-\beta,\cdots,-\beta\rangle_{1\cdots N}),
\end{eqnarray}
where $\beta =-i(\alpha /\sqrt{N})e^{-i\omega \pi /2},$ and the normalization
constant ${\cal N}^{\prime}=[2+2\cos \theta \exp (-2|\beta
|^2)]^{-1/2}$. Now the cavity field is in a vacuum state, and all its energy
has been transferred to the excitons.

Multipartite entanglement for the entangled coherent exciton state $|\Psi\rangle_{\text{excitons}} $ can be studied using the methods of the last section.
From Eqs.\,(\ref{eq:bbbbb}), (\ref{eq:fffff}), and (\ref{eq:tau4}) we obtain the quantities ${\cal B}$, ${\cal F}$, and the square of the multiqubit concurrence  for our state $|\Psi \rangle _{\text{excitons}}$ as

\begin{eqnarray}
{\cal B}&=&2^{N/2}(1-e^{-4|\alpha |^2/N})^{N/2},\\
{\cal F}&=&[1+(1-e^{-4|\alpha |^2/N})^{N/2}]^2/4+e^{-4|\alpha|^2}/4,\\
\tau _{1,\ldots ,N}&=&\frac{(1-e^{-4|\alpha |^2/N})^N}{(1+\cos \theta
e^{-2|\alpha |^2})^2}.
\end{eqnarray}

In Fig.\,3 we plot the quantity $B$, $F$, and $\tau$ against $|\alpha|$. 
We observe that the state is multipartite entangled when $|\alpha|$ is large enough. Quantitively the corresponding Bell inequality is violated when $|\alpha|>1.601$, and the state preparation fidelity ${\cal F}$ is larger than 1/2 when $|\alpha|>1.228$. We also observe that the square of the $(N+1)$-qubit concurrence is significantly larger than zero only if $|\alpha|$ is large enough. For fixed $N$ and very large  $|\alpha |$ the square of the $(N+1)$-qubit concurrence $\tau _{1,\ldots
,N}\approx 1$ which implies that the entangled coherent exciton states
becomes a GHZ-like state. On the other hand  $\tau _{1,\ldots ,N}\approx 0$
for fixed $N$ and small enough  $|\alpha |^2.$ As discussed in Ref.\cite
{WangBarry} the state $|\Psi \rangle _{\text{excitons}}$ with $\theta =\pi $
reduces to the multiqubit W state\cite{W} in the limit of $|\alpha |^2\rightarrow 0.$ It means that we can also prepare the W state in our system at time $t=\pi /2$
with the initial cavity field in a Fock state with one photon and all excitons
initially in the vacuum states. For the case of only two excitons the W state is just the maximally entangled state (one Bell state) as discussed by Liu {\it et al}.\cite{Yuxi}.

\section{Effects of environment on multipartite entanglement}

Environmental losses and decoherence are important effects in quantum information processing\cite{a1}. The lifetime of both cavity photon and exciton is generally considered to be on the order of picoseconds. Whereas, if we assume $\hbar g=0.5$ meV \cite{Yuxi}, the time we need to get maximal entanglement in our model is also on the order of picoseconds.  However, dynamical evolution suffers from decay of photons and excitons. Recent experiments show that the lifetime of photons and excitons can be prolonged\cite{Boyer,butov,Walther}.
In particular, for cavity decay, recent experiments display an elongated decay time of photons in the microwave domain\cite {Walther}. Thus, we only consider the decay of excitons in the following discussions.
 
We follow the method of Ref.\,\cite{Yuxi}, and assume that the environment is at zero temperature and the system dissipates by interaction between excitons and a multimode electromagnetic field. Under the rotating-wave approximation we write the Hamitonian as
\begin{eqnarray}
H&=&\omega b_0^\dagger b_0 +\omega \sum_{n=1}^N b_n^\dagger b_n+\sum_k \omega_k a_k^\dagger a_k \nonumber\\
&+&g\sum_{n=1}^N (b_n^\dagger b_0 +b_0^\dagger b_n)\nonumber\\
&+& \sum_{n=1}^{N}\sum_k
\lambda_k (b_n^\dagger a_k + b_n 
a^\dagger_k),
\end{eqnarray}
where $a_k^\dagger$ ($a_k$) denotes the creation (annihilation) operator of the multimode magnetic field with frequency $\omega_k$. We assume $g_1=g_2=\cdots=g$ implying that the excitons are equally coupled to the cavity mode, and $n$-independence of the $\lambda_k$ implying that the excitons are also equally coupled to the environment. 

Now we use the Heisenberg picture to study the problem. The Heisenberg equations 
for related operators are obtained as follows:
\begin{eqnarray}
\partial b_0/\partial t &=&i \omega b_0 +ig\sum_{n=1}^N b_n, \nonumber\\
\partial b_n/\partial t &=&i \omega b_n +ig b_0 +i\sum_k \lambda_k a_k, \nonumber\\
\partial a_k/\partial t  &=&i\omega_k a_k +i\lambda_k \sum_{n=1}^N b_n.
\end{eqnarray}
Note that we use a slightly different Heisenberg picture for our purpose to obtain the final state vector at time $t$. We let an operator evolve as $A(t)=\exp(-iHt)A(0)\exp(iHt)$ and it satisfies $i\partial A(t)/\partial t=[H,A(t)]$.

Now we introduce the collective exciton operator $b_c=1/\sqrt{N}\sum_{n=1}^N b_n$ which can be considered as a boson since $[b_c,b_c^\dagger]=1$. In terms of the collective boson operator we rewrite the above equation as
\begin{eqnarray}
\partial b_0/\partial t &=&i \omega b_0 +ig_N b_c, \nonumber\\
\partial b_c/\partial t &=&i \omega b_c +ig_N b_0 +i\sum_k \lambda_{N,k} a_k, \nonumber\\
\partial a_k/\partial t  &=&i\omega_k a_k +i\lambda_{N,k} b_c,\label{bbb}
\end{eqnarray}
where $g_N=g\sqrt{N}$ and $\lambda_{N,k}=\lambda_k\sqrt{N}$. If we let
\begin{equation}
b_0=B_0e^{i\omega t},\; b_c=B_ce^{i\omega t},\; a_k=A_k e^{i\omega t}, 
\end{equation}
Eq.\,(\ref{bbb}) reduces to
\begin{eqnarray}
\partial B_0/\partial t &=&ig_N B_c, \nonumber\\
\partial B_c/\partial t &=&ig_N B_0 +i\sum_k \lambda_{N,k} A_k, \nonumber\\
\partial A_k/\partial t  &=&i(\omega_k-\omega) A_k +i\lambda_{N,k}B_c.\label{bbbb}
\end{eqnarray}
To solve the above equation we make the Laplace transform
\begin{equation}
\bar{f}(s)=\int_{0}^\infty e^{-st}f(t) dt.
\end{equation}
We obtain after minor algebra
\begin{eqnarray}
s\bar{B}_0&=&B_0(0)+ig_N\bar{B}_c,\\
\bar{B}_c&=&\frac{B_c(0)+ig_N\bar{B}_0+i\sum_k\frac{\lambda_{N,k}A_k(0)}{s+i(\omega_k-\omega)}}{s+N\sum_k\frac{\lambda^2_k}{s+i(\omega_k-\omega)}}.
\end{eqnarray}
The above equation cannot be solved exactly. So we resort to the Wigner-Weisskopff approximation\cite{WW}. After the approximation, from the above two equations we obtain
\begin{equation}
\bar{B}_0=\frac{(s+\frac{N\Gamma}{2})B_0(0)+ig\sqrt{N}B_c(0)-\sum_k\frac{g_N\lambda_{N,k}A_k(0)}{s+i(\omega_k-\omega)}}{s^2+\frac{N\Gamma}{2} s+Ng^2},\label{barb}
\end{equation}
where $\Gamma=2\pi \epsilon(\omega)\lambda^2(\omega)$ and $\epsilon(\omega)$ is a distribution function of the multimode electromagnetic field. We assumed that 
$\Delta\omega=-\int d\omega_k \epsilon(\omega_k)\lambda^2(\omega_k)/(\omega_k-\omega)=0$ in the derivation of the above equation.

From Eq.\,(\ref{barb}) we obtain the operator $b_0(t)$ in the Heisenberg representation
\begin{eqnarray}
b_0(t)&=&u(t)b_0(0)+v(t)\sum_{n=1}^Nb_n(0)+\sum_{k}w_k(t)a_k(0),\nonumber\\
u(t)&=&e^{-N\Gamma t/4}[\cos(\Delta_N t)+\frac{N\Gamma}{4\Delta_N}\sin(\Delta_Nt)]e^{i\omega t},\nonumber\\
v(t)&=& \frac{ig}{\Delta_N}e^{-N\Gamma t/4}\sin(\Delta_N t)e^{i\omega t},\label{buv}
\end{eqnarray}
where $\Delta_N=\sqrt{Ng^2-N^2\Gamma^2/16}$. 
Let the cavity field be an odd coherent state and other systems in the vacuum states. Then, from Eq.\,(\ref{buv}) we obtain the state vector at time $t$ as
\begin{eqnarray}
|\Psi(t)\rangle&=&[2-2\exp(-2|\alpha|^2)]\nonumber\\
&&\times (
|\alpha_u\rangle\otimes |\alpha_v\rangle^{\otimes N}\otimes \prod_{k}|\alpha _{w,k}\rangle \nonumber\\
&&-|-\alpha_u\rangle\otimes |-\alpha_v\rangle^{\otimes N}\otimes \prod_{k}|-\alpha_{w,k}\rangle),    \label{fff1}   
\end{eqnarray}
where $\alpha_u=\alpha u^*(t)$, $\alpha_v=\alpha v^*(t)$, 
and $\alpha_{w,k}=\alpha w^*_k(t)$. 

We use the state preparation fidelity ${\cal F}$ to examine multipartite entanglement in the above state and choose the following GHZ state for consideration,
\begin{equation}
|\psi_{\text{GHZ}}\rangle=\frac{1}{\sqrt{2}}(
|\alpha_u\rangle\otimes|\alpha_v\rangle^{\otimes N}-
|\alpha_u^\perp\rangle\otimes|\alpha_v^\perp\rangle^{\otimes N}),\label{fff2}
\end{equation}
where 
\begin{eqnarray}
&&|\alpha_x^\perp\rangle=(|-\alpha_x\rangle-p_x|\alpha_x\rangle)/{\cal M}_x,\nonumber\\
&&p_x=\exp(-2|\alpha_x|^2), \nonumber\\
&&{\cal M}_x=\sqrt{1-p_x^2}, \; x=u,v.
\end{eqnarray}
{}From Eqs.\,(\ref{fff1}) and (\ref{fff2}) we obtain the state preparation fidelity as
\begin{eqnarray}
{\cal F}&=&\langle\Psi(t)|\psi_{\text{GHZ}}\rangle\langle\psi_{\text{GHZ}}|\Psi(t)\rangle\nonumber\\
&=&\frac{1}{4-4e^{-2|\alpha|^2}}[1+2e^{-2|\alpha|^2}({\cal M}_u{\cal M}_v^Np_u^{-1}p_v^{-N}-1)\nonumber\\
&&+({\cal M}_u{\cal M}_v^N-p_up_v^N)^2].
\end{eqnarray}
In the derivation of the above equation we have used the relation
\begin{equation}
|\alpha|^2=|\alpha_u|^2+N|\alpha_v|^2+\sum_k |\alpha_{w,k}|^2,
\end{equation}
which results from the normalization of the state $|\Psi(t)\rangle$  and implies 
the energy conservation.

Figure 4 shows the fidelity against time $t$ for different numbers of excitons. 
{}From the figure we see that the fidelity is not a periodic function of $t$ due to the dissipation of  energy to the environment. When there is no dissipation ($\Gamma=0$), the fidelity attains the ideal case of being a periodic function of $t$. We observe that multipartite entanglement occurs only in the beginning of the evolution. When the number of excitons becomes larger, the generation of the multipartite entanglement becomes more difficult. In the limit of $t\rightarrow \infty$ the fidelity becomes 0.5 as we expected. In this case there is no multipartite entanglement and all the energy of the cavity-excitons system dissipate to the environment. The environment diminishes the generation of multipartite entanglement when the number of excitons increases. From Eq.\,(\ref{buv}) we can see that the decay rate is proportional to the number of excitons when the intensity of the cavity field is fixed. In addition, we find a similar result as that of Ref.\,\cite{Yuxi} if we fix $\Gamma$ and vary $|\alpha|$, i.e., the multipartite entanglement decays rapidly with  increasing the cavity field density.

\section{Conclusions}

In conclusion, we have investigated the dynamical evolution of multipartite
entanglement in a system of quantum dots embedded in a microcavity. The
entanglement is studied via two sufficient conditions for multipartite entanglement and the square of the multiqubit concurrence. 
We observed the global
multipartite entanglement and at certain times the entanglement becomes
maximal. We can also produce the multipartite entangled coherent exciton
states and multiqubit W state by preparing different initial states. 
Finally, we study the effects of environment on the generation of multipartite entangled states, and find that the decay rate is proportional to the number of excitons. We also find that the entanglement decays rapidly with increasing the cavity field density.

Although multipartite entanglement studied here has not yet been observed experimentally, 
the potential application of excitons in quantum computing as well as rapid development of CQED technique suggest that our analysis will find applications in this field.

{\bf ACKNOWLEDGEMENTS}
We appreciate the helpful discussions with Paolo Zanardi,
Irene D'Amico, and Ehoud Pazy. This work has been supported by the European Commission through the Research Project SQUID within the FET Program and IST, Q-ACTA, and by an Australian Research Council Large Grant.

\begin{figure}
\begin{center}
\epsfig{width=10cm,file=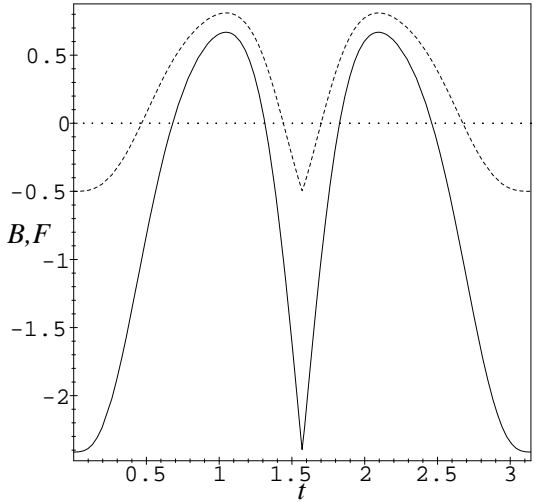}
\caption[]{The quantities $B$ (solid line) and $F$ (dashed line) against time. 
For $B$ we choose $\theta=\beta_N$. For $F$ we choose $\theta=\gamma=\pi/2$.
The parameter $G=1$, $|\alpha|^2=3$, and $N=3$.} 
\end{center}
\end{figure}

\begin{figure}
\begin{center}
\epsfig{width=10cm,file=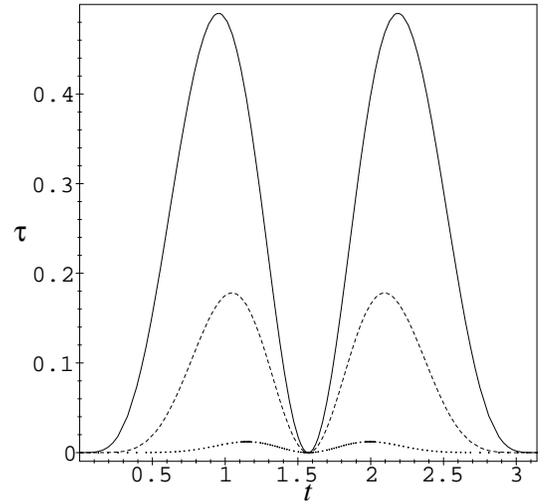}
\caption[]{The square of the multiqubit concurrence against time. The parameter $g=1$, $|\alpha|^2=0.9$, 
$\theta=\pi$, and all $g_i$ are equal. The solid, dashed and dotted lines correspond to 
$N=2$, 3 and 5 respectively.}
\end{center}
\end{figure}

\begin{figure}
\begin{center}
\epsfig{width=10cm,file=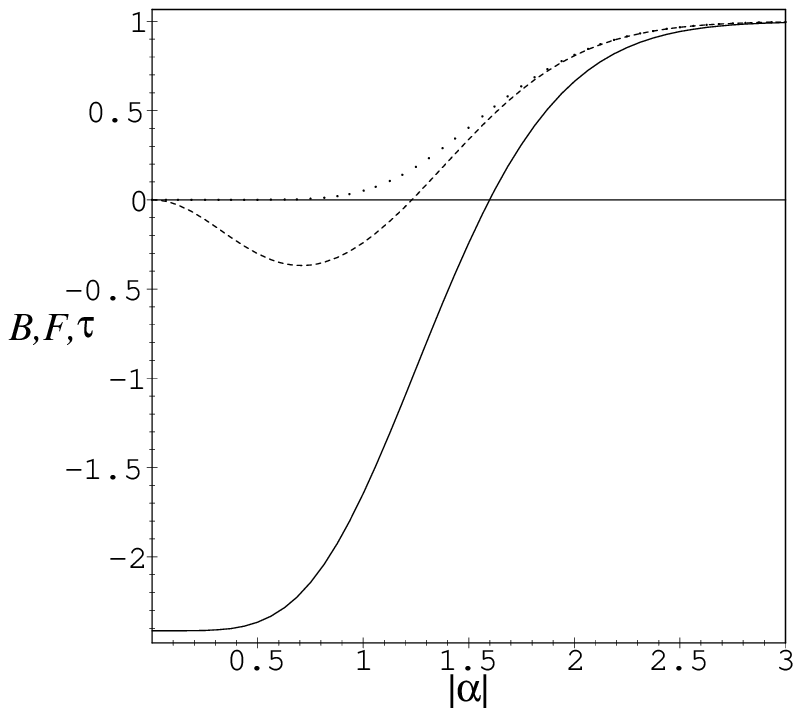}
\caption[]{The quantity $B$ (solid line), $F$ (dashed line), and square of the multiqubit concurrence (dotted line) against $|\alpha|$. The parameter $N=5$.}
\end{center}
\end{figure}

\begin{figure}
\begin{center}
\epsfig{width=10cm,file=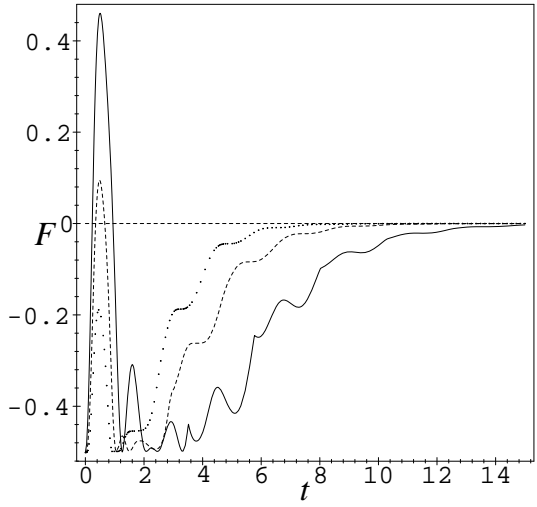}
\caption[]{The quantity $F=2{\cal F}-1$ for different numbers of excitons: $N=2$\,(solid line), $N=3$ (dashed line) and $N=4$ (dotted line). The parameters $|\alpha|^2=3, g=1 $, and $\Gamma=0.5$.}
\end{center}
\end{figure}

\end{document}